# Putting Together the Pieces:
# A Concept for Holistic Industrial Intrusion Detection[1]

Simon Duque Antón and Hans Dieter Schotten
German Research Center for Artificial Intelligence, Kaiserslautern, Germany
Simon.Duque_Anton@dfki.de
Hans_Dieter.Schotten@dfki.de

**Abstract:** The fourth industrial revolution, resulting in Industry 4.0, provides a variety of novel business cases. These business cases provide benefits with respect to cost, effort, customer satisfaction and production time. Progress in production can be monitored in real-time by the customer, maintenance can be performed in a remote fashion, time- and cost-efficient production of customer specific products is enabled. These business cases are founded on characteristics of digitisation, namely an increase in intercommunication and embedded computational capacities. Besides the advantages derived from the ever present communication properties, it increases the attack surface of a network as well. As industrial protocols and systems were not designed with security in mind, spectacular attacks on industrial systems occurred over the last years. Most industrial communication protocols do not provide means to ensure authentication or encryption. This means attackers with access to a network can read and write information. Originally not meant to be connected to public networks, the use cases of Industry 4.0 require interconnectivity, often through insecure public networks. This lead to an increasing interest in information security products for industrial applications. In this work, the concept for holistic intrusion detection methods in an industrial context is presented. It is based on different works considering several aspects of industrial environments and their capabilities to identify intrusions as an anomaly in network or process data. These capabilities are based on preceding experiments on real and synthetic data. In order to justify the concept, an overview of potential and actual attack vectors and attacks on industrial systems is provided. It is shown that different aspects of industrial facilities, e.g. office IT, shop floor OT, firewalled connections to customers and partners are analysed as well as the different layers of the automation pyramid require different methods to detect attacks. Additionally, the singular steps of an attack on industrial applications are characterised. Finally, a resulting concept for integration of these methods is proposed, providing the means to detect the different stages of an attack by different means

## 1. Introduction

The fourth digital revolution, creating Industry 4.0, enables a plethora of novel use and business cases for industrial applications. Founded on the development in communication and computation technology, the Industry 4.0 use cases employ the increase of intercommunication and embedded intelligence, e.g. the digital factory (Stef et al. 2013). These novel business cases can decrease cost, time to production and effort. They enable customer individual production, direct information about the state of production, remote maintenance and operation. Since they rely heavily on communication, connection through network boundaries is essential. However, industrial communication protocols have not been designed with security in mind. Most protocols, such as *Modbus* (Modbus 2012; Modbus-IDA 2006) or *Profinet* (PROFIBUS 2017) do not contain methods to authenticate entities or encrypt communication. Originally, industrial control systems, commonly known as Supervisory Control And Data Acquisition (SCADA) or Operation Technology (OT)–Systems were meant to be physically separated from public networks. Additionally, highly application specific devices and configurations were considered to make it exceedingly difficult for an attacker to exploit the systems (Igure et al. 2006). These assumptions, however, have proven to be wrong. A spectacular series of successful attacks on industrial

---

companies and systems shows that industry has become a target of cyber criminals and state-sponsored actors alike (Duque Anton et al. 2017a). Commercial Off The Shelf (COTS)–products make integration and configuration of new devices easier for operators. On the other hand, they allow attackers to analyse and develop exploits against the devices. Furthermore, the network layout of industry is based on well-established protocols and network segmentation techniques. The application of common techniques makes it easy for attackers to reuse attack techniques. The incentive for attackers is manifold. Attacks on industrial and critical infrastructure facilities can originate in a political agenda, even though attribution is difficult. Different industrial attacks in the past, e.g. the Ukrainian blackout in December 2015 and *Stuxnet*, are rumoured to be part of a political plan. Additionally, cybercrime has become a profitable business. In 2015, the revenue of cyber attacks has surpassed the global profit in drug trafficking for the first time (Leyden 2018). The potential of cybercrime in industry, e.g. by sabotage and espionage, is deemed high.

Industrial applications are characterised by unique properties. They commonly contain a network segment of classic office Information Technology (IT) that is used for communication with customers and partners as well as raw material and product configuration, so-called Enterprise Resource Planning (ERP) and Manufacturing Execution Systems (MESs). Additionally, there is an OT network segment used for control of the production machines. The networks are usually separated by data diodes or De-Militarized Zones (DMZs). Cyber attacks aim on both of these network segments, depending on their goal. However, these kinds of networks are vastly different in terms of connected devices, protocol and usage behaviour. Thus, detecting and preventing attacks on industrial facilities is strongly application- and attack-specific. A thorough understanding of the domain and possible attacks is required, as well as an understanding of promising methods to detect attacks in the individual domains. In this work, an overview of possible attacks on industry is provided, including original vectors and methods for lateral movement. Based on this, promising intrusion detection methods are proposed. Their applicability is derived from prior experiments. Due to the specific nature of industrial environments, different parts of the network and different attack vectors need different methods to detect them. Thus, the concept for a hybrid model integrating different methods required for detecting typical attacks along their path is proposed.

The remainder of this work is structured as follows. In Section 2, an overview of related work is provided. A discussion of attacks on industrial systems can be found in Section 3. Approaches to detect such attacks at different points of the system are introduced in Section 4. A hybrid industrial intrusion detection approach, consisting of a combination of the introduced methods, is presented in Section 5. This work closes with a discussion provided in Section 6.

## 2. Related Work

In this section, works addressing attacks on industrial settings are discussed at first. After that, works discussing industrial intrusion detection are presented and categorised according to aim and method of detection.

There is a variety of scientific discussions on risks and threats in SCADA-systems, e.g. *Zhu et al.* (2011) and (Igure et al. 2006). In their work, they discuss the inherent weaknesses of SCADA systems, such as lack in authentication and encryption. *Virvillis and Gritzalis* address the effects of Advanced Persistent Threats (APTs) on industry using the examples of four widely discussed industrial malwares: *Stuxnet*, *Duqu*, *Flame* and *Red Ocotber* (Virvilis and Gritzalis 2013). *Positive Technologies* provide an exhaustive analysis of the state of industrial and corporate network security as well as risks and threats on industry (Positive Technologies 2018). Additionally, the individual attacks have been discussed: *Langner* provides a thorough analysis of the *Stuxnet* attacks (Langner 2013). *Lindsay* discusses *Stuxnet* while putting it in the context of cyber warfare and describing its limits (Lindsay 2013). *Lee et al*, *Cherepanov* and *Dragos* discuss the industrial malware *Industroyer,* also known as *Crashoverride* (Lee et al. 2016; Cherepanov 2017; Dragos 2016). It is responsible for the successful attacks on the Ukrainian power grid in December 2015. *Shamir* analyses a new version of *BlackEnergy* that is linked to cyber attacks on the Ukrainian government as well as the aforementioned blackouts (Shamir 2016).

In order to address these issues and to protect industrial networks from attacks, a variety of research has been conducted. Due to their periodic, repetitive nature, state information can be used to detect anomalies. This is done by *Khalili and Sami* (2015). *Caselli et al.* focus on sequences in a way that intrusion detection systems are able to understand the step in a process chain they are currently in (Caselli et al. 2015). Thus, attacks based on reaching unwanted, but generally allowed states, can be detected. The same approach is chosen by *Fovino et al.* with a focus on *Modbus* and *DNP3* (Fovino et al. 2010). The determinism found in industrial applications is made use of by *Hadeli et al.* (Hadeli et al.). A similar approach is chosen by *Morris et al.*, they present a system that can be used to detect anomalies in *Modbus RTU* and ASCII-systems in a retrofit fashion (Morris et al. 2012). *Gao and Morris* discuss potential attacks on industrial systems and intrusion detection-based countermeasures for *Modbus*-based communication (Gao and Morris 2014). *Ponomarev and Atkison* made use of telemetry information to obtain insights about threats and attacks (Ponomarev and Atkison 2016).

The increase in computational power allows more complex algorithms, such as methods of Artificial Intelligence (AI) or Machine Learning (ML), to work in a time-efficient manner. An overview of the challenges is provided by *Mantere et al.* (2012). *Borges Hink et al.* aim at differentiating between disturbances based on natural deviations and attacks in power grids in order to support human decision making (Borges Hink et al.). *Beaver et al.* evaluate ML-based methods for detecting malicious SCADA communication (Beaver et al. 2013). Furthermore, *One-Class Support Vector Machine (SVM)* as well as *Ant Colony Optimisation* are used to detect anomalies in industrial networks (Shang et al. 2015; Tsang and Kwong).

Wireless technologies are getting used in an increasing fashion for industrial applications. They are easy and quick in set up and operation. This motivates research on securing wireless channels, e.g. Mobile Ad hoc NETworks (MANETs) (Shakshuki et al. 2013), Wireless Sensor Networks (Shin et al. 2010) and wireless industrial networks (Wei and Kim 2012).

## 3. Attacks on Industry – An Overview

Attacks on industrial environments are different from attacks on classic home- and office-IT systems. Industrial companies employ IT as well as OT networks. An exemplary network structure found in industrial environments is shown in Figure 1.

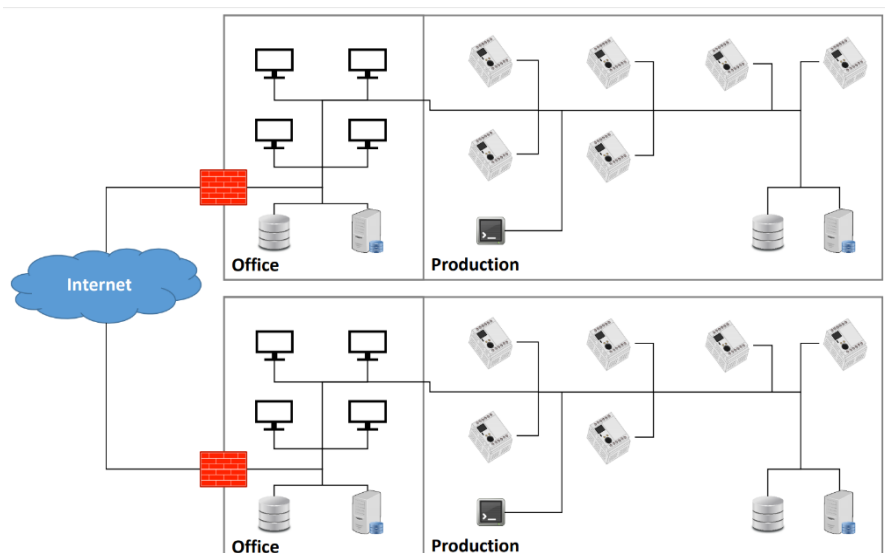

Figure 1: Typical Structure of Industrial Networks

It can be seen that an office-IT network is connected to the public Internet. This is used for customer contact, company representation and other tasks. However, this network is also used for configuration, creation and modification of OT-level applications, e.g. code for Programmable Logic Controllers (PLCs). If an attacker aims at sabotaging applications, this is the point to execute the payload. If the network where a PLC control is created is connected to a public network, no air gap is in action, allowing for an attacker to compromise OT devices as soon as the IT-perimeter is broken. According to *Positive Technologies*, an attacker could have penetrated the network and accessed the corporate information system in 73 % of the cases they evaluated (Positive Technologies 2018). Many attacks on industrial applications were possible due to phishing, an exemplary set of attacks is shown in Figure 2.

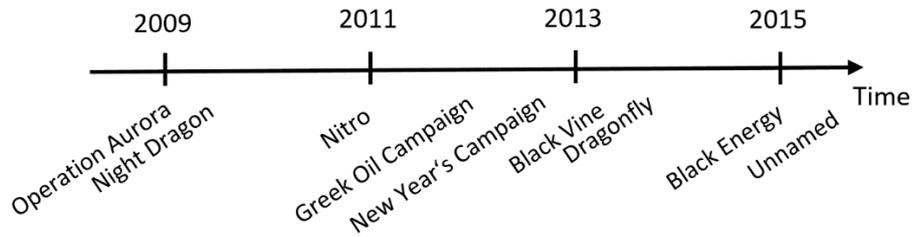

**Figure 2: Overview of Industrial Phishing Attacks**

*Operation Aurora* targeted *Google* and other software companies (McClure et al. 2010). *Night Dragon*, *Greek Oil* and *New Year's* campaign targeted the energy sector, especially petrol processing facilities (Wueest 2014), while *Nitro* targeted chemical processing facilities (Chien and O'Gorman 2011). *Black Vine* targeted aerospace and healthcare companies (DiMaggio 2015). *Dragonfly* (Symantec Corporation 2014) and *Black Energy* were specifically aimed at Industrial Control Systems (ICSs), with *Black Energy* (Lee et al. 2016) being linked to the Ukrainian power grid failure in December 2015. The *Unnamed* campaign was used to extract information from ICS systems, making it a corporate espionage tool (Kaspersky Lab 2017).

Apart from spear phishing, any means to break the perimeter allows an attacker leverage to move laterally inside the network. USB flash drive- or SD memory card-based attacks for bridging air gaps have been established, e.g. by the *Stuxnet* attack (Langner 2013). Detecting attacks once the malware has moved to the OT is difficult for several reasons, they are listed in Table 1.

**Table 1: Comparison of IT and OT**

| Property | IT | OT |
|---|---|---|
| *Operation times* | Commonly around 3 to 5 years | 2 to 4 decades |
| *Cost of interruption* | Medium | High, hundreds of thousands per hour |
| *Placement & Management* | Central, in similar location | Spatially distributed |
| *Upgrade capabilities* | Upgrades possible | Upgrades difficult |
| *Software properties* | Similar, compatible solutions | Proprietary, vendor-specific solutions |
| *Cost* | Moderate | High |
| *Effects* | Only digital domain | Physical world |

This table shows the difficulty in securing industrial facilities. Their cost and operation time makes add-on solutions for security necessary. However, they need to cope with difficulties to be distributed and upgraded due to software management and distribution of machines. Furthermore, a plethora of proprietary protocols needs to be addressed while downtimes are not acceptable.

## 4. Industrial Intrusion Detection Approaches

A generic example of a typical industrial network structure was presented in Section 3, pictured in Figure 1. This structure indicates possible places in the system to detect attacks at different stages. In this section, all of those places are discussed. At first, detection is possible when the outer perimeter is broken. After that, lateral movement within the industrial IT network is the goal. Once the malware reached the OT network, it can be detected either by suspicious network traffic or by anomalous process behaviour.

### 4.1 Detecting Perimeter Breaks

The task to detect breaches of the perimeter is located in the area of classical IT security, it is not in the scope of this work. It should be noted, however, that the human factor is considered to be the most significant risk with respect to breaches of the perimeter. Phishing and other forms of social engineering have proven to create a significant effect on employees. If a single employee falls for a scam e-mail, attackers can obtain a foothold within the network. For tricking certain, usually high privileged, employees, more targeted methods, such as spear phishing with plausible information or water holing attacks, offer promising results. Unfortunately, technical means can only provide a minor amount of protection against social engineering. Awareness and training of responsible personnel is necessary, in combination with a sound but useable security concept (Surveillance Self-Defense 2018).

## 4.2 Detecting Lateral Movement

Similar to the detection of perimeter breaches, lateral movement in the IT-network need to be detected by well-established IT-security tools, such as Network- and Host-based Intrusion Detection and Prevention Systems (NIDS, HIDS, NIPS, HIPS). Such tools, e.g. firewalls, detect and mitigate the propagation of malware, often with the aid of sophisticated heuristics or Deep Packet Investigation (DPI). Furthermore, the usage of Security Information and Event Management (SIEM) systems can be useful for detecting attacks. Such systems, e.g. *Splunk* (2019) often provide means for the integration of process and OT data as well, allowing industrial IT-security personnel to obtain holistic information about the security status of a network or system.

## 4.3 Detecting Suspicious OT Traffic

In contrast to IT-networks, traffic in OT-networks is usually using proprietary protocols that do not consider security objectives. However, the traffic is much more periodic and repetitive than in IT networks. This property can be used to detect attacks in forms of anomalies. Most commercial industrial intrusion detection systems employ a baselining-algorithm that monitors the network for a certain amount of time as a training. After that, deviations from the baseline are flagged as anomalous. Additionally, scientific approaches have been evaluated in order to find anomalies. *Matrix Profiles*, as presented by Yeh et al. (2016), have proven to be efficient in detecting anomalies in OT-network traffic (Duque Anton et al. 2018a). This property is shown in Figure 3.

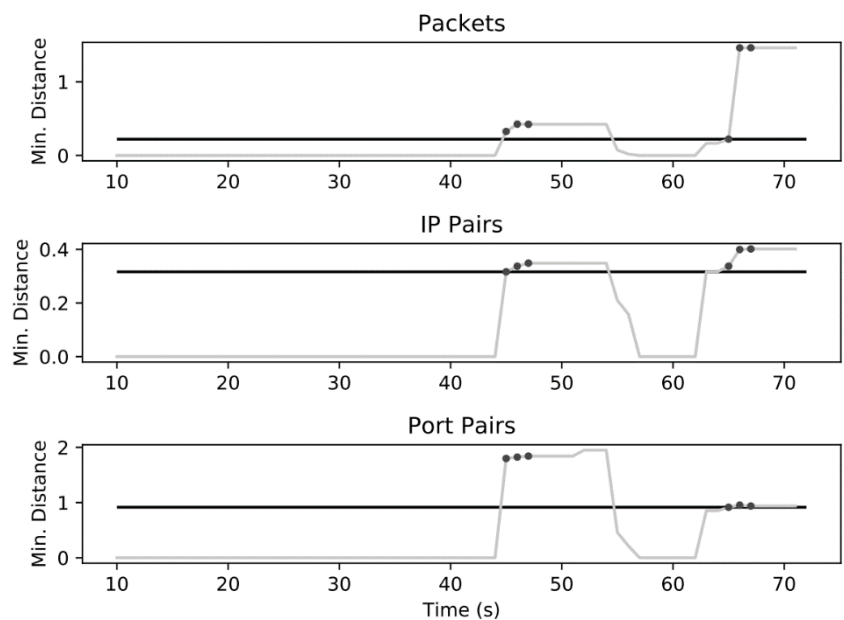

**Figure 3: Example of *Matrix Profiles* to Detect Malicious OT traffic**

*Matrix Profiles* are calculated based on a distance metric, e.g. the z-Normlized Euclidean distance as follows:

$$d(\hat{x}, \hat{y}) = \sqrt{2m(1 - \frac{\sum_{i=1}^{m} x_i\, y_i - m\mu_x\mu_y}{m\sigma_x\sigma_y})}$$

The black dots indicate the attacks, the grey line indicates the minimal distance, and the black line indicates the minimum threshold recognising all attacks. *Matrix Profiles* are a windowed approach where for each sequence of length *m* the minimal distance to all other sequences of length *m* is calculated. *M* is the only hyper parameter that needs to be set by the user. Furthermore, *Matrix Profiles* do not need training as such, as they are applied directly to the data. If the minimal distance of a given sequence to all other sequences is small, it is a common motif, meaning it occurs often, indicating normal behaviour. If the minimal distance is high, however, the motif is a singularity which indicates an outlier as well as a potential attack. The evaluation in Figure 3 is based on an industrial data set introduced by *Lemay and Fernandez* (2016). They created an emulation of an electric circuit breaker system communicating via *Modbus*, consisting of three to twelve Remote Terminal Units (RTUs) and one to two Master Terminal Units (MTUs). In doing so, several data sets were monitored, some containing simulated human interaction as well as periodic polling. After that, different kinds of attacks were introduced to a set of the data sets. The attacks are based on TCP/IP, e.g. scanning the network, extracting information or uploading malicious files. The data set under analysis is called "CnC_uploading_exe_modbus_6RTU_with_operate". As features, the numbers of packets, IP- and port-pairs were used. They provide valid indicators of intrusions.

In addition to the time series analysis of the OT network traffic, packet-based analysis can be used to produce promising results as well. *Random Forests*, *Support Vector Machines* (*SVM*), *k-means clustering* and *k nearest neighbours* have been evaluated on three of the data sets provided by *Lemay and Fernandez* (Duque Anton et al. 2018c). Since *SVM* and *Random Forests* provide promising results, the application of both of them on the data sets "Moving_two_files_Modbus_6RTU" (DS1), "Send_a_fake_command_Modbus_6RTU_with_operate" (DS2) and a combination of data sets with and without attacks (DS3). This shows packet-based analysis is a promising approach as well. The concepts of *SVM* and *Random Forest are shown in Figure 4.*

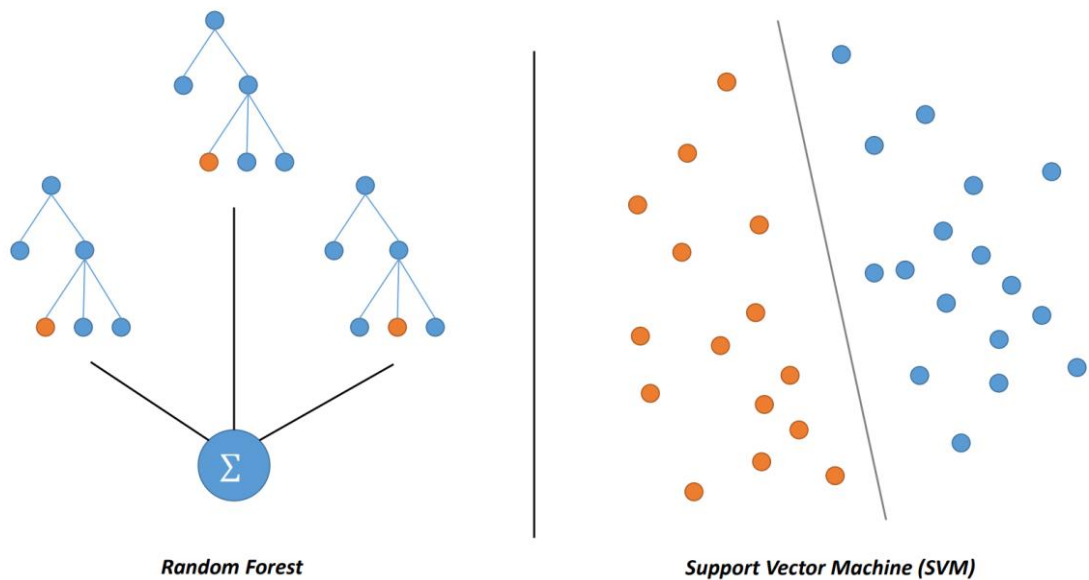

**Figure 4: Random Forest and SVM**

*Random Forests* consist of a group of decision trees of which the majority decision is used as the classification. *SVM* is a large-margin classifier, it aims at separating clusters such that the distance of each entity from the separator is maximal.

**Table 2: Performance of *SVM* and *Random Forest* on OT Network Traffic**

|          | SVM      |          | Random Forest |          |
|----------|----------|----------|---------------|----------|
| Data Set | F1-score | Accuracy | F1-score      | Accuracy |
| DS1      | 1.0      | 1.0      | 1.0           | 1.0      |
| DS2      | 1.0      | 1.0      | 0.999851      | 0.999701 |
| DS3      | 0.999968 | 0.999936 | 0.999986      | 0.999973 |

## 4.4  Detecting Suspicious Process Behaviour

In addition to meta-information, time series approaches such as *Matrix Profiles*, can also be applied to process information, as shown in Figure 5. In the first and third line, the water flow and water tank level of an industrial batch process are shown, as explained in (Duque Anton et al. 2019b) and (Duque Anton et al. 2019a). The second and fourth row describe the according minimal distance. Attacks are introduced at a packet count of 4,000 and 6,500 as analysed and explained by (Duque Anton et al. 2019b). This corresponds to the notable increases in minimal distance, successfully detecting the disruptions of the process as an anomaly.

## 5. A Hybrid Concept for Industrial Intrusion Detection

In order to detect the attacks as described in Section 3 as thoroughly as possible, the methods presented in Section 4 need to be combined in an intelligent matter. On the one hand, strong IT-security is required to detect and mitigate intrusions in and lateral movement through the IT network. Additionally, a combination of OT-network and process-based intrusion detection can be used to detect and attribute attacks. If host-based information, e.g. from HMIs, is combined with OT-network data, a dashboard as exemplarily shown in Figure 6 can aid a human operator in detecting attacks, obtaining an overview of the network state and being able to place attacks at their points of origin and destination, thus helping to provide efficient Incident Response (IR). Especially in the OT-domain, time series-based anomaly detection has shown potential, making it a plausible candidate for an intrusion detection mechanism (Duque Anton et al. 2018a).

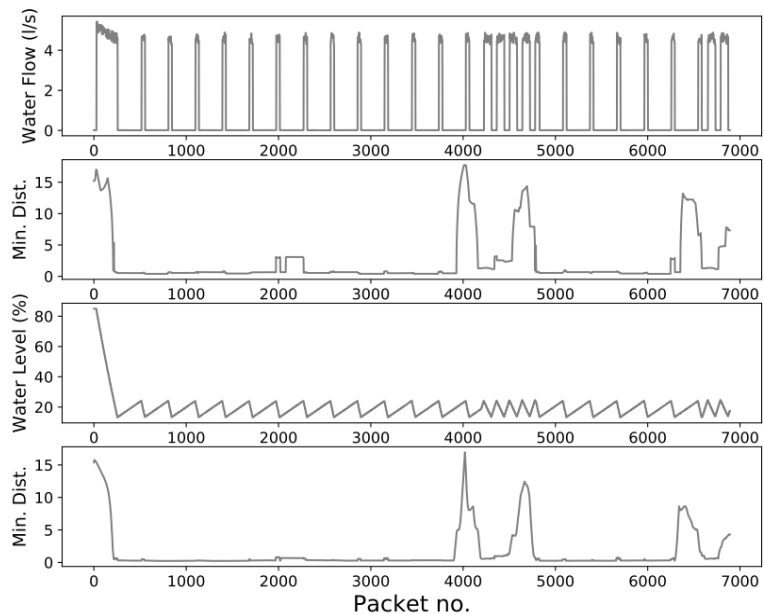

**Figure 5: Example of *Matrix Profiles* to Detect Malicious Process Behaviour**

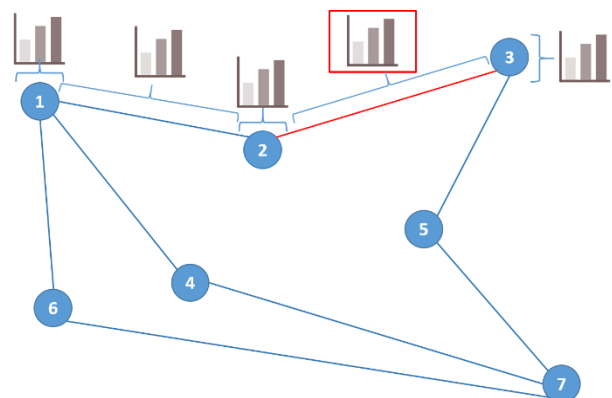

**Figure 6: Concept for a Hybrid Industrial IDS**

Combining IT-, OT- and process information has the potential to detect attacks during at least one stage. First, the IT-network has to be breached and traversed. If this happens successfully, OT- or process information can still be used to detect artefacts or effects of an attack, thus increasing the probability of noting an attack.

Additionally to the singular information sources, combining information sources and considering contextual information can provide crucial insights into the security status (Duque Anton et al. 2017b; Duque Anton et al. 2018b).

## 6. Conclusion and Outlook

In this work, an overview of industrial attack vectors was provided. The stages an attacker needs to overcome in order to successfully penetrate an industrial network, i.e. breaking the perimeter, moving laterally, exploiting industrial hardware, executing the attack, becoming permanent, have been discussed. After that, possible places in the networks commonly found in an industrial environment are presented. It shows that time series are especially promising for the OT and process domain. Finally, the combination of these approaches is expected to address an attack throughout the complete attack cycle, providing a holistic view and aiding the detection.

## 7. Acknowledgements

This work has been supported by the Federal Ministry of Education and Research (BMBF) of the Federal Republic of Germany within the project IUNO Insec (KIS4ITS0001). The authors alone are responsible for the content of the paper.

## 8. Literaturverzeichnis